\documentclass[aps,prl,twocolumn,preprintnumbers,amsmath,amssymb]{revtex4-1}
\usepackage{graphicx}
\usepackage{subfigure}
\usepackage{epsfig}
\usepackage{dcolumn}
\usepackage{bm}
\usepackage{ulem}
\usepackage{color}
\usepackage{float}

\def\abs#1{\left|#1\right|}

\def\be{\begin{equation}}       \def\ee{\end{equation}}
\def\bea{\begin{eqnarray}}      \def\eea{\end{eqnarray}}
\def\ba{\begin{array}}
\def\ea{\end{array}}
\def\bnum{\begin{enumerate} }
\def\enum{\end{enumerate}}

\def\nn{\nonumber}

\def\=>{\Rightarrow}
\def\>{\rightarrow}

\def\PRB{Phys. Rev. B}
\def\PRL{Phys. Rev. Lett.}
\def\eye2{Fathbb{I}}

\def\Tr{\mathrm{Tr}}

\renewcommand{\>}{\rangle}

\newcommand {\C}{\textcolor {red}}

\newcommand{\al}[1]{\begin{align}#1\end{align}}
\newcommand{\eq}[2]{
	\begin{equation}
	#1 \label{#2}
	\end{equation}
}
\newcommand{\md}{\mathrm{d}}
\newcommand{\re}[1]{\frac{1}{#1}}
\newcommand{\me}{\mathrm{e}}
\newcommand{\mi}{\mathrm{i}}

\newcommand{\dpf}[2]{\dfrac{\partial #1}{\partial #2}}

\newcommand{\ddf}[2]{\dfrac{\mathrm{d} #1}{\mathrm{d} #2}}
\newcommand{\vect}[1]{\boldsymbol{#1}}

\newcommand{\ipic}[4]{
	\begin{figure}[H]\centering
		\includegraphics[#4]{#1}
		\caption{#2}  \label{#3}
	\end{figure}
}

\newcommand{\itpic}[9]{
	\begin{figure}\centering
		\begin{tabular}{cc}
			\begin{minipage} [t]{#9\textwidth}
				\centering
				\includegraphics[#4]{#1}
				\caption{#2}  \label{#3}
			\end{minipage}
			\begin{minipage}[t]{#9\textwidth}
				\centering
				\includegraphics[#8]{#5}
				\caption{#6}  \label{#7}
			\end{minipage}
		\end{tabular}
	\end{figure}
}

\newcommand{\fkh}[1]{\left[ #1 \right] }

\begin{document}
\title{Correlated triple-Weyl semimetals with Coulomb interactions}
\author{Shi-Xin Zhang}
\author{Shao-Kai Jian}
\author{Hong Yao}
\email{yaohong@tsinghua.edu.cn}
\affiliation{Institute for Advanced Study, Tsinghua University, Beijing 100084, China
}

\begin{abstract}
We study interaction effects, including both long-ranged Coulomb and short-range interactions, in three-dimensional topological triple-Weyl semimetals whose triple-Weyl points are protected by crystal symmetries. By performing Wilsonian renormalization group analysis of the low-energy effective field theory of the minimal model with triple-Weyl nodes, we find that the fixed point of noninteracting triple-Weyl fermions is unstable in the presence of Coulomb interactions and flows to a nontrivial stable fixed point representing marginal Fermi liquids with anisotropic screening effects. We further discuss relevant unusual physical consequences due to the novel behavior of correlation effects in this system.
\end{abstract}

\date{\today}
\maketitle

\textbf{Introduction:}
Topology has played a central role in modern condensed matter physics\cite{Fradkinbook,XGWenbook,volovikbook}. It is sometimes surprisingly useful and profound to characterize band structures or phases of matters via topology\cite{TKNN,haldane1983}. In the past years, topological insulators and superconductors have been extensively  studied\cite{kane2005, bernevig2006, fu2007, Hasan-10,XLQi-11,moore2010,schnyder2008,ryu2010}. More recently, topological semimetals have been of great interest\cite{XGWan-11,GXu-11,Burkov2011,ZWang-12,Rappe12,nagaosanc,hosur2013,SMHuang-15,HWeng-15, Soluyanov2015,Ryu,Ruan2015,Ruan2016}, especially after the experimental observation of Weyl fermions\cite{Weyl} in TaAs family\cite{SYXu-15,BQLv-15,BQLv-15b, LXYang-15, xu2015b} and type-II Weyl fermions in MoTe$_2$\cite{shuyunzhou,Kaminski,Chen1604,XJZhou,Shi}. 
Weyl points, protected by the lattice translational and particle number conservation symmetries, are crossing points
of two non-degenerate bands around which the dispersion is linear\cite{nielsen1981a, nielsen1981b, NN1983}.
Such Weyl nodes act like a monopole of Berry curvature in crystal momentum space with monopole charge $\pm 1$.

Besides the usual Weyl fermions with monopole charge $\pm 1$ described above, there are other types of Weyl fermions when point-group symmetries are further taken into considerations. For instance, double (triple)-Weyl fermions, with quadratic (cubic) dispersions in two momentum directions and linear dispersion in the third direction, are protected by rotation symmetries of certain point groups 
and further classified according to $ C_n $ rotation symmetry\cite{CFang-12}. Double (triple)-Weyl points can be treated as magnetic monopoles of Berry curvature in crystal momentum space with twice (three times) the topological charge of usual single-Weyl points and can split into two (three) single-Weyl points without point group symmetry protections. Since the largest rotation symmetry in solid is $ C_6 $, it was pointed out that triple-Weyl fermion is the one with the highest monopole ($\pm 3$) charges 
amongst the family of Weyl fermions with two-fold degeneracy that can be protected by symmetries. Although features of single- and double-Weyl fermions have been extensively studied \cite{GXu-11,SMHuang-15,SKJian-15,HHLai-15, jian2016a}, properties of the new type of Weyl semimetals, triple-Weyl fermions with monopole charges $\pm 3$, remain largely unknown. In this paper, we study the correlation effect in triple-Weyl fermions to partially fill this gap. 

Correlation effect is one of central issues in condensed matter physics. Interaction effects in electronic systems with only discrete Fermi points can be very different from systems with large Fermi surface\cite{Shankar}, which have attracted lots of interest recently considering both long-ranged Coulomb interactions\cite{Goswami-11,Isobe-12,Isobe-13} and short-ranged interactions\cite{Herbut-06,Herbut-09,KaiSun-09,QingLiu-10,Joseph-14,Savary-14,roy2016}. Systems with only several discrete Fermi points show various novel correlated effects such as non-Fermi liquid behavior\cite{Abrikosov-74,Moon-13}, topological Mott insulator phase\cite{Herbut-14,Janssen-15}, anisotropic screening of Coulomb interactions\cite{Abrikosov-72,BJYang-14,SKJian-15,HHLai-15, isobe2016} or even emergent space-time supersymmetry in the infrared limit\cite{SSLee-07,KunYang-10,Grover-14,SKJian-14,jian2016b,ponte2014}. For triple-Weyl semimetals, one of typical electronic systems with only Fermi points, effects of long-ranged Coulomb interactions need be considered carefully due to the vanishing densities of states on Weyl points where Thomas-Fermi screening mechanism fails. Therefore, it is natural to ask whether long-ranged Coulomb interactions can induce novel physical consequences or exotic phases in triple-Weyl semimetals.

Here we consider the simplest model for triple-Weyl semimetals with only two triple-Weyl nodes which is the minimum number of Weyl points according to the no-go theorem\cite{nielsen1981a, nielsen1981b, NN1983}. From renormalization group (RG) analysis, we find that at the noninteracting Gaussian fixed point corresponding to free triple-Weyl fermions, the anisotropy of Coulomb interaction is relevant. As a result, the noninteracting fixed point is unstable and the system flows to a stable fixed point with anisotropic Coulomb potential where the strength of interaction is  marginal irrelevant. The cubic-dispersing directions of triple-Weyl fermions have larger density of states in low energy, which results in stronger screening effect than the linear-dispersing direction. Specifically, Coulomb interaction is screened to a faster $\frac1{r^3}$ decaying behavior along cubic-dispersing directions compare to unchanged $\frac{1}{r}$ decay along the linear-dispersing direction.

We further discuss unusual physical consequences due to exotic behavior of Coulomb interactions in triple-Weyl semimetals, which are closely related to experiments. For instance, we obtain an anisotropic screening of a charge impurity in  noninteracting triple-Weyl semimetals as well as in triple-Weyl semimetals with Coulomb interactions where the renormalized Coulomb potential should be taken into account. Moreover, owing to the marginally irrelevance of Coulomb interactions at the stable fixed point, the system behaves as a marginal Fermi liquid, where various physics observables receive logarithmical corrections.   We specifically calculate the temperature dependence of specific heat and find exotic logarithmical corrections explicitly,  which can be measured in future experiments after the relevant materials are discovered.

\textbf{Model and theory:} We start with deriving the effective Hamiltonian for triple-Weyl fermions by symmetry analysis. Based on the previous analysis in Ref. \cite{CFang-12}, we know that the $C_6$ rotation symmetry is necessary to protect triple-Weyl fermions in solid state materials. To write down the symmetry-allowed Hamiltonian, for simplicity, we consider the largest point-symmetry group, $ C_{6h} $, that contains $C_6$ symmetry.  By symmetry arguments (we leave detailed symmetry analysis in Appendix), we have the Hamiltonian for free triple-Weyl fermions as
\bea\label{hamiltonian}
H_f
\!=\!v_1(k_x^3-3k_x k_y^2)\tau_3\sigma_1\!+\!v_1(k_y^3-3k_y k_x^2)\tau_3\sigma_2\!+\!v_3 k_z\tau_3\sigma_3,\nn\\
\eea
where $ \tau $ and $ \sigma $ are Pauli matrix for valley ($ \pm K_z$) and spin ($ S_z=\pm\frac{3}{2} $ subspace of $ S=\frac{3}{2} $) degrees of freedom, $v_1$ is the ``velocity'' in the $k_x$$k_y$ plane, and $v_3$ is the Fermi velocity along the $k_z$ axis.

By the Hubbard-Stratonovich transformation, we introduce a static bosonic field to decouple long range four-fermion Coulomb interactions such that the effective action can be written as: 
\bea
S &=& S_f + S_b + S_{bf}, \\
S_f &=& \int \md^3 k \md \omega \; \psi^\dagger_{\vect{k}} \Big[-i\omega
+ H_f(\vect k) \Big] \psi_{\vect{k}}, \\
S_b &=& \int \md^3 k \md \omega~ \re{2}\phi_{-\vect k} [k_x^2+k_y^2+\eta k_z^2] \phi_{\vect k}, \\
S_{bf} &=& \mi g \int \md^4 x\; \phi\psi^\dagger\psi ,\label{action}
\eea
 where $\psi^\dag=(\psi^\dag_+,\psi^\dag_-)$ is a four component spinor with two-component $\psi^\dag_\pm$ represents the low-energy electrons near two band-touching points $\pm K_z$; $\eta$ is a real parameter parameterizing the anisotropy of Coulomb interactions; and $g$ is the coupling strength. 
Note that we have neglected the couplings between different triple-Weyl fermions as they are irrelevant when the two triple-Weyl points are finitely separated in momentum space.

\textbf{ RPA analysis:} We now implement a random phase approximation (RPA) calculation to study the screening effect with  triple-Weyl fermions. 
Starting with low-energy actions given by Eq. \eqref{action}, we integrate out fermions (see the Appendix for details). The result shows that electron-hole polarization is anisotropic as
\bea
    -\Pi(\vect p) & \propto& p_\perp^2+|p_z|^{\frac{2}{3}}, \label{rpa}
\eea
where $p_\perp\!=\! \sqrt{p_x^2+p_y^2}$, and we have neglected insignificant coefficients as well as higher order terms. The screened Coulomb potentials are then given by  
\eq{V(\vect p)\propto\re{p_x^2+p_y^2+p_z^2-\Pi(p)}\approx\re{p_x^2+p_y^2+|p_z|^{\frac{2}{3}}},}{trv} where we omit the qudratic term of $ p_z $ which is less important than $  p_z^{2/3} $ part when momentum transfer is small.

The anisotropic momentum dependence in Eq. \eqref{trv} means that the bare isotropic Coulomb interaction becomes anisotropic because of the screening by the low-energy anisotropic triple-Weyl fermions. By performing Fourier transformations, the form of Coulomb potential along the $z$ axis is unchanged as  $V(0,z) \propto \frac{1}{|z|}$, while in $xy$ plane, the Coulomb potential decays faster as $V(r_\perp,0) \propto \frac{1}{\abs{r_\perp}^3}$ due to the stronger screening effects of the cubic-dispersing fermions in the $xy$ plane.

\textbf{Renormalization group analysis:} We now perform a Wilsonian momentum-shell renormalization group (RG) analysis to study the effect of Coulomb interactions in triple-Weyl fermions and see if Coulomb interactions could drive the system into a nontrivial non-Fermi liquid phase without well-defined quasiparticles but with finite interaction strength in the fixed point. 

 The tree level scaling dimensions of spatial coordinates are different for different directions as a result of the anisotropic dispersion for triple-Weyl fermions.  Therefore, we can assume that scaling dimensions are  $[k_z]=z_3,[k_x,k_y]=z_1,[\omega]=1$.  For simplicity, we further assume $ v_1=v_3=1 $ below, since their renormalization effects of them can be absorbed into $z_1$ and $z_3$.

\begin{figure} \centering   \subfigure[ ] { \label{RG}    \includegraphics[width=0.42\columnwidth]{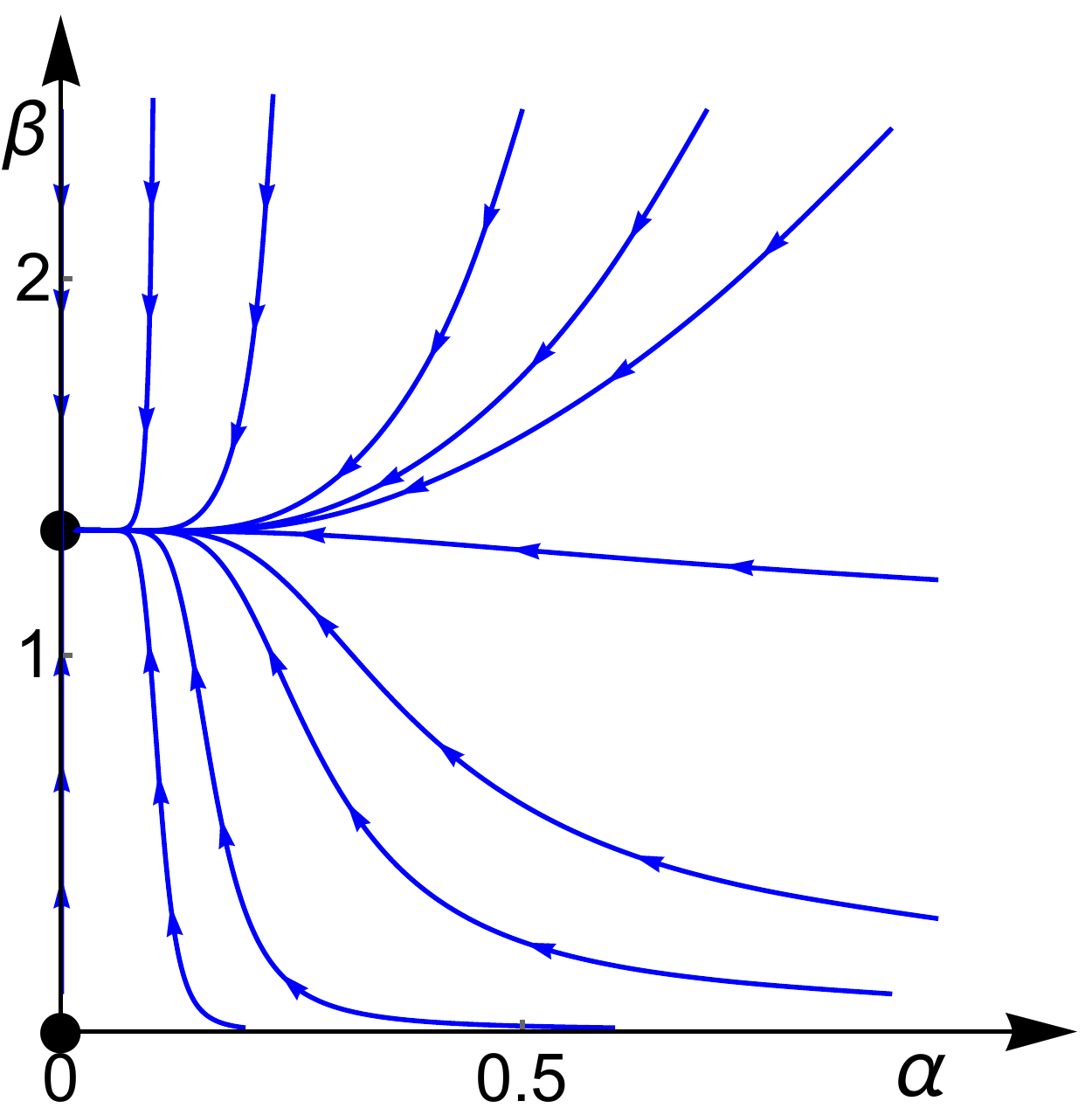} }    \subfigure[ ] { \label{IC}    \includegraphics[width=0.42\columnwidth]{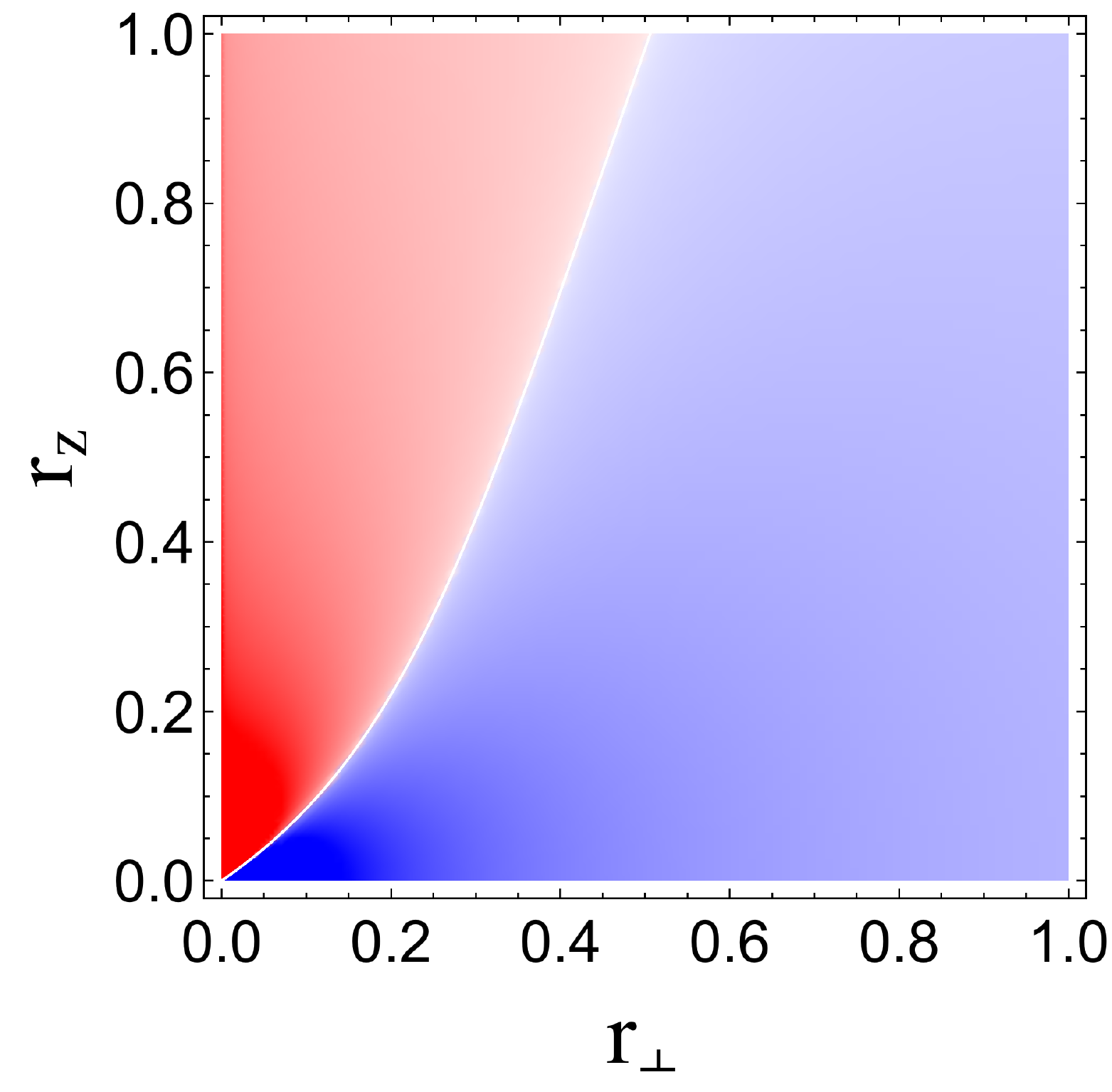}    }    \caption{ a: RG flow diagram in $ \alpha, \beta $ parameter space: where $ \alpha\equiv g^2 $, $ \beta\equiv\frac{g^2}{6\pi^2 Q^4\eta} $. Two fixed points are shown in this flow diagram. The non-interacting Gaussian fixed point $ (0,0) $ is unstable against Coulomb interactions and flows to $ (0,4/3) $, which is the stable fixed point  with $ \eta=\alpha/\beta\rightarrow 0 $, a sign of anisotropic Coulomb screening while the interaction strength $ \alpha $ is still marginally irrelevant in the stable fixed point.\\
			b:  Real space distribution of induced charge by a charge impurity at origin in the simplest model with only two triple-Weyl nodes in $ k_z $ axis in the non-interacting limit: Blue color for positive charge while red color for negative charge with darker sides stand for larger magnitude in densities of charges. }    \label{fig}    \end{figure}

While the full RG flow equations are summarized in the Appendix, here we outline the main results of RG analysis. 
For convenience, we combine the coupling constants in a more compact way by defining $\alpha\equiv g^2$, $\beta\equiv\frac{g^2}{6\pi^2 Q^4\eta}$, where $Q$ is the momentum cutoff of the cylinder geometry in low-energy effective actions. By  iterating the RG equations numerically, we can obtain the RG flow of $\alpha$ and $\beta$, as shown in Fig. \ref{RG}. There are only two fixed points: $(\alpha^*,\beta^*)=(0,0)$ is the unstable fixed point characterizing the noninteracting triple-Weyl fermions and $(\alpha^*,\beta^*)=(0,\frac{4}{3})$ is the stable one representing a new marginal Fermi liquid phase with anisotropic screening effects. Near the fixed points, the linearized RG equations  of $\alpha$ and $\beta$ are given by
\bea 
\dpf{\alpha}{l}&\approx& -0.18\alpha^2, \label{rgeq1}\\ \dpf{\beta}{l}&\approx&\beta(\frac{4}{3}-\beta+0.026\alpha).\label{rgeq2}
\eea
 From RG equations, we find the Coulomb interaction strength  $\alpha$ is marginally irrelevant at the stable fixed point and Coulomb interaction becomes strongly anisotropic, namely, $\eta\propto \alpha/\beta \to 0$ in the stable fixed point which is consistent with our RG results by directly using $ g $ and $ \eta $ as flowing parameters. (Note that $\eta=1$ corresponds to isotropic Coulomb interaction.) Furthermore, at the nontrivial fixed point, i.e., $\alpha=0$, the scaling properties of spatial coordinates remain the same as the tree-level results, namely, $z_1=\frac{1}{3},z_3=1$.

This anisotropic character of the screened Coulomb interactions can be derived from the particle-hole polarization,
   \eq{\vect{p}^2-\Pi(\vect{p})=p_\perp^2(1+\frac{3\alpha}{2\pi^2}l)+\eta p_z^2(1+\beta l),}{}
   where terms beyond unity are given by corrections from one-loop Feynman diagrams.
At the stable fixed point $(\alpha^*,\beta^*)=(0,\frac{4}{3})$,  boson's kinetic energy in the $p_xp_y$ plane gets no renormalization because $ \alpha=0 $, while it is  renormalized along the $p_z$ axis, i.e. 
\eq{\vect{p}^2-\Pi(\vect{p})=p_\parallel^2+ p_z^2(1+\frac{4}{3}\ln \frac{\Lambda}{v |p_z|})\sim p_x^2+p_y^2+|p_z|^{\frac{2}{3}},}{}
where $\Lambda$ is the energy cutoff.  The renormalized Coulomb potential hence behaves as $V(\vect p)=\frac{1}{p_x^2+p_y^2+|p_z|^{2/3}}$. This result is consistent with the RPA calculations in Eq. \eqref{rpa}, which further confirms our conclusion on the strong anisotropic screening of the Coulomb interactions.

\textbf{Screening of a charged impurity:}
The unusual behaviors of Coulomb interactions give rise to various novel physical consequences. Here we investigate the distribution of charges induced by a charge impurity. Suppose one puts an impurity at the origin with electric charge $ Ze $. By linear response theory, the induced charge distribution is given by $ \rho_{\textrm{ind}}(\vect{q})=ZeV(\vect{q})\Pi(\vect{q})$ \cite{kotov2008},  where $ V(\vect q)=\frac{g_0^2}{q^2} $ is the bare Coulomb potential for noninteracting case as well as $ V(\vect{q})=\frac{g^2}{q_x^2+q_y^2+|q_z|^{2/3}} $ for interacting case. Using Eq. (\ref{rpa})
we numerically evaluate the screening charge distribution in real space, as shown in Fig.\ref{IC}. 
One can see that the induced charge distribution is highly anisotropic due to unusual polarization. Taking account of the renormalization effects by gapless triple-Weyl fermions in the low energy limit, one should get $ \delta(r) $ distribution for induced charge\cite{BJYang-14} by using renormalized Coulomb potentials in Eq. (\ref{trv}).

\textbf{Logarithmic corrections to physical quantities:} Since the Coulomb interaction is marginally irrelevant at the stable fixed point, we expect marginal Fermi liquid behaviors which bring logarithmic corrections to  physical quantities.  
 Namely, considering the flow of interaction strength, according to Eq. (\ref{rgeq1}),
\eq{
	\alpha(E)= \frac{\alpha_0}{1+0.18\alpha_0 \ln \frac{\Lambda}{E}},}{fine_structure
}
where $\alpha_0$ is the bare parameter at energy scale $ \Lambda $. As the energy or equivalently temperature decreases, the interaction strength becomes weaker logarithmically which shall bring corresponding corrections to physical quantities compared to their original values in non-interacting cases.

Here, we calculate logarithmic corrections of specific heat for an example. According to $ C(T)\!=\!-T\frac{\partial^2 f}{\partial T^2} $ and the scaling dimension of free energy density, we get an approximate flow equation for specific heat based on dimension analysis: $\frac{\partial C}{\partial l}=C(2z_1+z_3)$, where $z_1$, $z_3$ are functions of $\alpha$  representing scaling dimensions of spatial coordinates(see more detail in Appendix). By flow equation of specific heat,  we can show that
\al{C&\propto T^{\frac{5}{3}}\Big(\frac{1+c_1\alpha_0 \ln T_0}{1+c_1\alpha_0 \ln T}\Big)^{-c_2}\notag\\&\times\me^{c_3[(\ln(\frac{\alpha_0}{1+c_1 \alpha_0\ln T}))^2-(\ln(\frac{\alpha_0}{1+c_1\alpha_0\ln T_0}))]^2}\label{second},}
where $ c_1=0.18$, $c_2=0.133$, $c_3=0.022 $ are some positive 
 constants. Note that the presence of the second line 
makes it decay faster than the usual logarithmic correction as the first line.

\textbf{Discussions and conclusion:}
So far, all calculations are based on  the minimal model of triple-Weyl semimetals with only two triple-Weyl nodes. It is worth noting that there can be at most one $ C_6 $ axis in a crystal lattice system. As a consequence, there is no material with multiple pairs of triple-Weyl fermions lying in different axes in crystal momentum space. This is in contrast with the case of double-Weyl semimetals for which multiple pairs of double-Weyl nodes are possible. Namely our simple model Eq. (\ref{hamiltonian}) describing triple-Weyl fermions already captures the main characters of triple-Weyl semimetals in realistic materials.

In order to clearly observe those exotic physics of Coulomb interactions in experiments, apart from two triple-Weyl nodes exactly at the Fermi level, there should be no coexisting trivial Fermi surface in the bulk. 
The materials hosting these symmetry-related triple-Weyl points without coexisting trivial Fermi surface are called ideal triple-Weyl semimetals. A number of candidate materials for ideal Weyl semimetals have been proposed (see, e.g. Ref. \cite{Ruan2015,Ruan2016}). It would also be interesting to search for candidate materials for ideal triple-Weyl semimetals.

In the end, we briefly discuss effects of the short-ranged interactions in such system. Short-range interactions can be written as four-fermion interactions, i.e. $ g_i(\psi^\dagger M\psi)(\psi^\dagger M'\psi) $, where $ M $ and $ M' $ are 4$\times$4 Hermitian matrices. Since the scaling dimension for $ g_i $ is $ -\frac{2}{3} $ at the tree level, four-fermion interactions are irrelevant near non-interacting Gaussian fixed point as well as the nontrivial stable fixed point with anisotropic Coulomb interactions. We expect our results are robust when short-range interactions are small. An interesting further work is to explore the fate of triple-Weyl fermions when short-ranged interactions become moderate or strong. 

To conclude, we have explored the correlation effect of the long-range Coulomb interactions in triple-Weyl semimetals. While the Coulomb interaction is marginally irrelevant at low energy, it becomes strongly anisotropic at the stable fixed point which represents a new phase with marginal Fermi liquid properties. Owing to those novel features of the interplay between Coulomb interactions and triple-Weyl fermions, many physical quantities receive logarithmic corrections in low temperature, which can be observed in ideal triple-Weyl semimetals after materials hosting these triple-Weyl fermions are discovered.

{\it Acknowledgement}:  This work is supported in part by the NSFC under Grant No. 11474175 at Tsinghua University (SXZ, SKJ, and HY) and by the National
Thousand-Young-Talents Program (HY).

\begin{widetext}
\section{Appendix}
\renewcommand{\theequation}{A\arabic{equation}}
\setcounter{equation}{0}
\renewcommand{\thefigure}{A\arabic{figure}}
\setcounter{figure}{0}

\subsection{1. Symmetry analysis on triple-Weyl fermions}
The four-components wave function to describe the simplest triple-Weyl fermion model can be summarized as
\eq{(\uparrow +,\downarrow +, \uparrow -, \downarrow -),}{lab}
where $ \uparrow,\downarrow $ represent the subspace of spin $ \frac{3}{2} $ which consists $ S_z=\pm\frac{3}{2} $ while $ S_z=\pm\re{2} $ has been projected out. And $ \pm $ represent the two valleys $ k=\pm K_z $ which we assume lie on z axis.

For each operator on that basis, a symmetry operation gives
\eq{O'(\vec{k})=U(T)^{-1}O(T^{-1}(\vec{k}))U(T).}{lab}

We can write down possible symmetry operation generators
\al{&T_1=(z,\frac{\pi}{3})~~U(T)=-\mi\tau_0\sigma_z~T^{-1}(\vec{k})=(\frac{k_x}{2}+\frac{\sqrt{3}k_y}{2},-\frac{\sqrt{3}k_x}{2}+\frac{k_y}{2},k_z)\\&T_2=(x,\pi)~~U(T)=\mi\tau_x\sigma_x~~~T^{-1}(\vec{k})=(k_x,-k_y,-k_z)\\&
	T_3=(y,\pi)~~U(T)=\mi\tau_x\sigma_y~~~T^{-1}(\vec{k})=(-k_x,k_y,-k_z)\\&T_4=(I)~~~~~U(T)=\tau_x\sigma_0~~~~T^{-1}(\vec{k})=(-k_x,-k_y,-k_z)\\&T_5=(M_z)~~U(T)=\mi\tau_x\sigma_z~~~~T^{-1}(\vec{k})=(k_x,k_y,-k_z)\\&T_6=(M_x)~~U(T)=\mi\tau_0\sigma_x~~~~T^{-1}(\vec{k})=(-k_x,k_y,k_z)\\&T_7=(M_y)~~U(T)=\mi\tau_0\sigma_y~~~~T^{-1}(\vec{k})=(k_x,-k_y,k_z)}
where $ T_1\sim T_7 $ are for $ C_6 $ rotation along $ z $ axis, $ C_2 $ rotation along $ x $ or $ y $ axis, space inversion, mirror reflection about $ xy $, $ yz $, $ xz $ planes.

Again, note that $ C_6 $ rotation symmetry is necessary to host stable triple-Weyl fermions, we consider all point groups in solids with $ C_6 $ symmetry and list them and their generators below.
\al{C_6:&~~~T_1\\C_{6v}:&~~~T_1,T_6,T_7\\C_{6h}:&~~~T_1,T_4,T_5\\D_6:&~~~T_1,T_2,T_3\\D_{6h}:&~~~T_1,T_2,T_3,T_4,T_5,T_6,T_7}

To host triple-Weyl fermions, we can see $ C_{6h} $ is the suitable point group with the maximum symmetry while symmetries in $ D_{6h} $ are too much and no symmetry allowed terms survive in Hamiltonian of triple-Weyl fermions.

Consider a general term in non-interacting triple-Weyl Hamiltonian as $ \psi^\dagger f(k) \tau_i\sigma_j \psi $, we can check the following terms are allowed by $ C_{6h} $ symmetry
\eq{(k^3_y-3k_x^2k_y )\tau_{2,3}\sigma_{1,2}~~~~(k^3_x-3k_y^2k_x )\tau_{2,3}\sigma_{1,2}~~~k_z\tau_{2,3}\sigma_3.}{free}
So  free triple-Weyl fermion Hamiltonian can be chosen as
\eq{H_f=v_1(k_x^3-3k_y^2k_x)\tau_3\sigma_1+v_1(k^3_y-3k_x^2k_y )\tau_3\sigma_2+v_3k_z\tau_3\sigma_3.}{lab}
Note this free Hamiltonian  is just one choice but not the unique one. Actually, this free Hamiltonian possesses larger symmetry than $ C_{6h} $. We assume coefficients before the first two terms in Hamiltonian to be equal, which is unnecessary for that the two terms are separately two one-dimensional representations of point group $ C_{6h} $. Anyway, there is no qualitative difference in the results no matter what value those coefficients are and the identical coefficient $ v_1 $ introduce an enlarged symmetry, continuous rotation symmetry along $ z $ axis, which is more suitable when we study short-ranged interactions in triple-Weyl semimetals later. We can easily check the monopole charges for the two triple-Weyl nodes are indeed $ \pm 3 $.

\subsection{2. Random phase approximation analysis on triple-Weyl fermions}

The screening effect of Coulomb interaction by the low-energy triple-Weyl fermions is captured by polarization,
\bea
	\Pi(p) &=& g^2 \int \frac{\md^4 k}{(2\pi)^4} \Tr[G(k) G(k+p)] \nonumber\\
	&=& 4g^2  \int \frac{\md^4 k}{(2\pi)^4} \frac{-w(w+\Omega)+v_1^2(d_1(k)d_1(k+p)+d_2(k)d_2(k+p))+v_3^2
    k_z(k_z+p_z)} {(w^2+v_1^2(d_1^2(k)+d_2^2(k))+v_3^2k_z^2)((w+\Omega)^2+v_1^2(d_1^2(k+p)+d_2^2(k+p))+v_3^2(k_z+p_z)^2)},~~~~~
\eea
where $p=(\omega,\vect{p})$, and $d_1(k)=k_x^3-3k_y^2k_x$, $d_2(k)=k_y^3-3k_x^2k_y$.
 We can check that $\Pi(\Omega=0,\vec p=0)=0$ which is consistent with the vanishing density of states at the triple-Weyl points.  We  calculate the above integral numerically with only $ p_z $ or $ p_\perp $ nonzero and use a cylinder geometry integral region in momentum space.

The results are
\eq{\Pi(0,0,p_z)\propto p_z^{\frac{2}{3}},~~~\Pi(p_\perp,0)\propto p_\perp^2.}{lab}
Relevant numerical fit plots are shown as Figure \ref{1} and Figure \ref{2}

\itpic{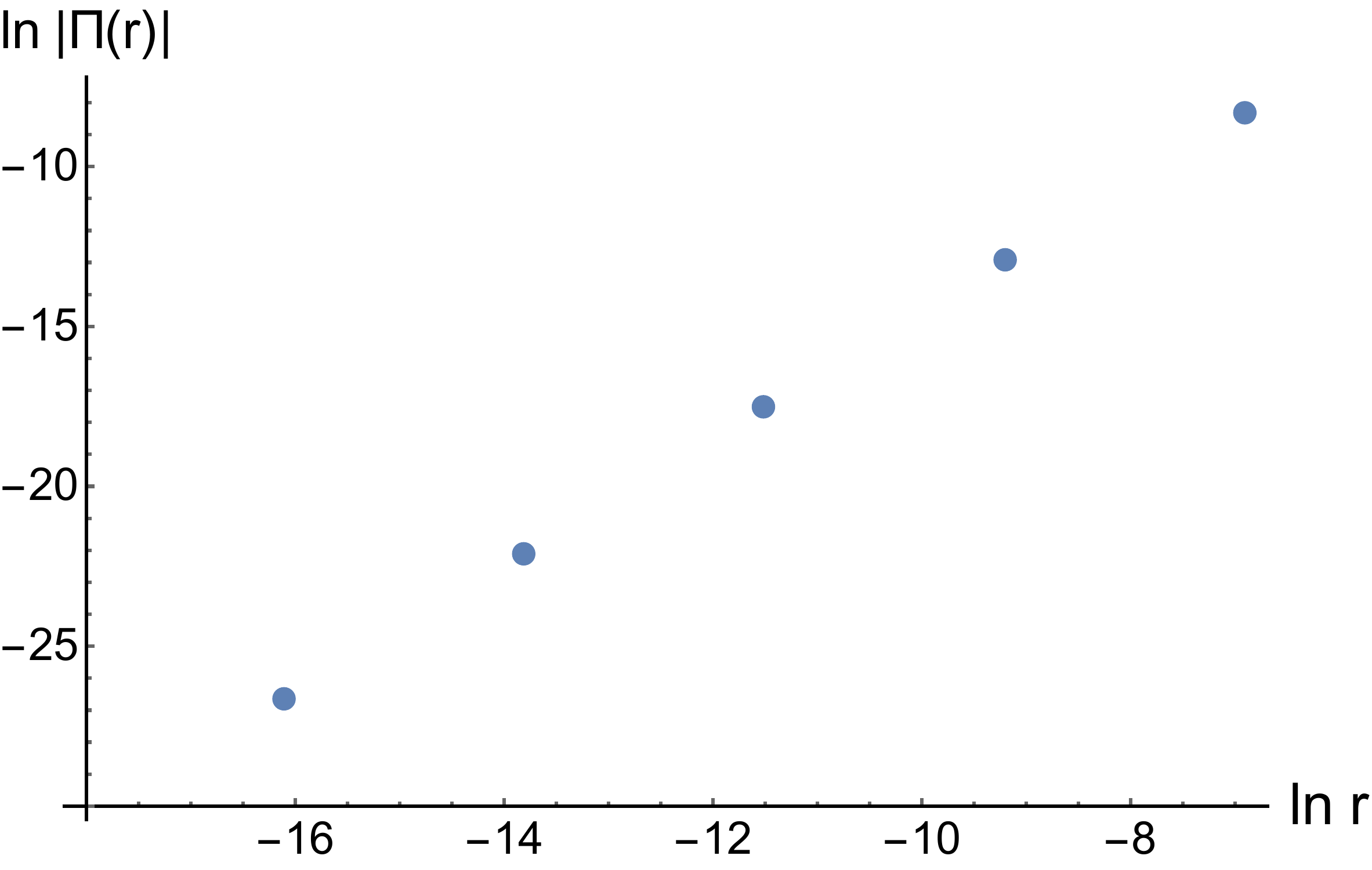}{The plot for polarization in plane (slope  $ k=2 $)}{1}{width=6.5cm}{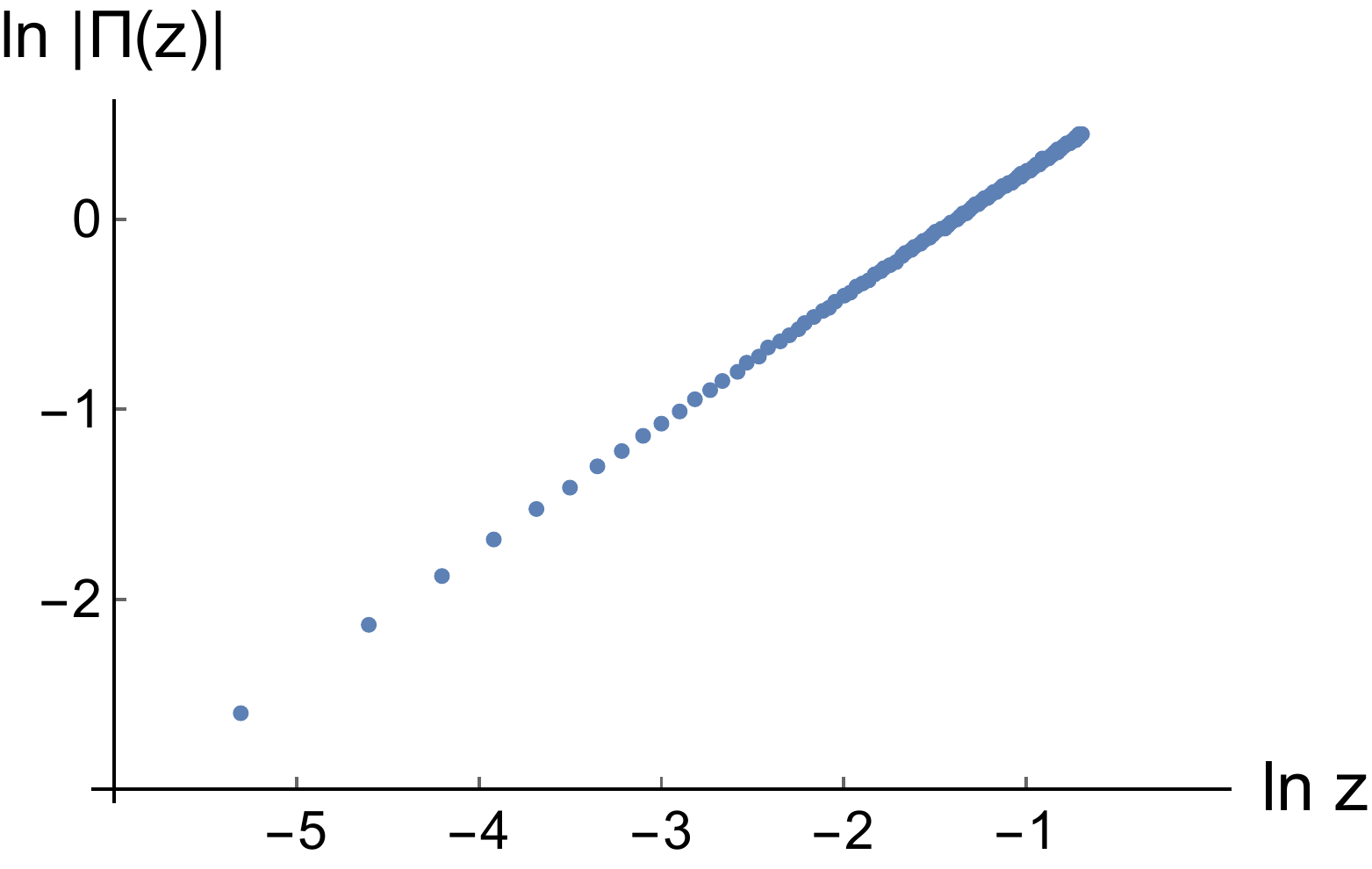}{The plot for polarization in $ z $ axis (slope  $ k=\frac{2}{3} $)}{2}{width=6.5cm}{0.5}
 Therefore, the full electron-hole polarization is approximated by:
	\begin{eqnarray}
	 \Pi(\vect{p})&\propto& -(p_\perp^2 +p_z^{\frac{2}{3}}).
	\end{eqnarray}
When performing Fourier transformations into real space, the Coulomb interaction along $z$ axis remain the same as the non-interacting limit: $V(0,0,z) \propto \frac{1}{|z|}$; while in the $xy$ plane, it is given by $V(x,y,0) \propto \frac{1}{(x^2+y^2)^{3/2}}$ which decays faster. The reason for the weaker or ``shorter'' ranged renormalized interaction in the $xy$ plane is because higher power law of dispersions  render larger density of states for fermions at low energy which screen the interaction more effectively.

\subsection{3. Renormalization group analysis on triple-Weyl fermions}
We give detailed renormalization group (RG) analysis in this section. The effective action for triple-Weyl fermions with  Coulomb interactions is summarized as
\al{S=&\int \md^3 k \md \omega \; \psi^\dagger(\mi\omega+v_1 (k_x^3-3k_x k_y^2)\tau_3\sigma_1+v_1(k_y^3-3k_y k_x^2)\tau_3\sigma_2+v_3 k_z\tau_3\sigma_3) \psi\notag\\&
+\int \md^3 k \md \omega \;	\re{2}\phi(k_x^2+k_y^2+\eta k_z^2)\phi+\mi g\int\md^3 k\md^3 q\md\omega\md\nu\;  \phi\psi^\dagger\psi,}
where the ultraviolet cutoff is implicit assumed which can be denote by energy cutoff $ \Lambda $ or equivalently by momentum cutoff $ Q $.

We  iteratively integrate momentum shells with momentum $Q_\perp\in (Q\me^{-l},Q)$ in a cylinder geometry, where $l$ is the RG flow parameter.  There are three relevant diagrams: boson polarization, fermion self-energy and vertex correction. At one-loop level,  fermions cannot get an anomalous dimension since the boson field is not dynamical and vertex correction vanishes due to Ward identity.
	
Two nonzero corrections at one-loop level  are
\eq{\Pi(k)=g^2\pi_1(v_1,v_3,Q)l(k_x^2+k_y^2)+g^2\pi_3(v_1,v_3,Q)l k_z^2,}{lab}
\al{\Sigma (k)=g^2 \xi_1 (v_1,v_3,\eta,Q)l ((k_x^3-3k_xk_y^2)\tau_3\sigma_1+(k_y^3-3k_x^2k_y)\tau_3\sigma_2)+g^2\xi_3(v_1,v_3,\eta,Q)lk_z\tau_3\sigma_3,}
where $ \pi_1,\pi_3 $,  $\xi_1,\xi_3$ are several well-defined  functions of $ v_1,v_3,\eta,Q $. Including the two corrections in original field theory, we have corrections on the action as
\al{\delta S=\int \md^3 k\md \omega \psi^\dagger(\delta v_1 l((k_x^3-3k_xk_y^2)\tau_3\sigma_1+(2k_y^3-6k_x^2k_y)\tau_3\sigma_2)+\delta v_3l k_z\tau_3\sigma_3)\psi+\re{2}\phi^2(\eta_\phi l(k_x^2+k_y^2)+\delta \eta l k_z^2),}
where we define terms as
\eq{\delta v_1=-g^2\xi_1,~~~\delta v_3=-g^2\xi_3,~~~\eta_\phi=-g^2\pi_1,~~~\delta \eta=-g^2\pi_3.}{lab}

	By scaling dimension analysis, we can set $ \fkh{\omega}=1,\fkh{k_{x,y}}=z_1,\fkh{k_z}=z_3 $ as mentioned in the main text and thus $ \fkh{\psi}=-z_1-z_3/2-1, \fkh{\phi}=-2z_1-z_3/2-1/2+\eta_\phi/2 $. Moreover $ \fkh{v_1}=1-3z_1 $, $ \fkh{v_3}=1-z_3 $, $\fkh{\eta}=2z_1-2z_3-\eta_\phi  $, $ \fkh{g}=-z_3/2+1/2-\eta_\phi/2 $.
	And RG flow equations are
	\al{\ddf{v_1}{l}&=(1-3z_1)v_1+\delta v_1,\\
\ddf{v_3}{l}&=(1-z_3)v_3+\delta v_3,\\
\ddf{\eta}{l}&=(2z_1-2z_3-\eta_\phi)\eta+\delta \eta,\\
\ddf{g}{l}&=(-z_3/2+1/2-\eta_\phi/2)g.}
To make physical quantities namely parameters in energy spectrum of fermions $ v_1,v_3 $ fixed, we  set
\eq{z_1=\frac{\delta v_1}{3v_1}+\re{3},~~~z_3=\frac{\delta v_3}{v_3}+1.}{lab}
and then we are left with two RG equations
\al{\ddf{\eta}{l}&=(-\frac{4}{3}+\frac{2g^2\xi_3}{v_3}-\frac{2g^2\xi_1}{3v_1}+g^2\pi_1)\eta-g^2\pi_3,\\
\ddf{g}{l}&=\frac{g^3}{2}(\frac{\xi_3}{v_3}+{\pi_1}).}

We can write down those functions explicitly:
\al{\pi_1&=-\frac{3}{2\pi^2 v_3},~~~~~~\pi_3=-\frac{v_3}{6\pi^2Q^4 v_1^2},\\
\xi_1&=\frac{v_1}{192\pi^2 v_3}\frac{\sqrt{1/x-1}x(44(x-1)x+15)-3(2x-1)(8(x-1)x+5)\arccos \sqrt{x}}{\sqrt{1/x-1}(x-1)^3},\\
\xi_3&=\frac{-1}{4\pi^2}\frac{x(-1+x+\sqrt{-1+1/x}\arccos \sqrt{x})}{(x-1)^2},}
where $ x= \frac{v_3^2}{v_1^2 \eta Q^4}$ and those four functions are always negative independent of arguments. Since $ v_1,v_3 $ doesn't affect the final results qualitatively, we set them to be unity for simplicity below. By this result, the beta function for $ g $ is always negative and only $ g=0 $ can be a solution, when $ g=0 $, beta function for $ \eta $ is zero when $ \eta=0 $, too. Thus $ ( g , \eta )=(0,0) $  is a stable fixed point with marginal irrelevant $ g $ and irrelevant $ \eta $ which is beyond Gaussian non-interacting one with  $ ( g, \eta )=(0,1) $.  Namely $ \eta $ always flow to zero even though the bare value is one which is a strong signal of anisotropic screening of Coulomb interactions. This is  consistent with our RPA calculations above.

To make the RG flow equations and formalism more clear, we introduce dimensionless variables as
\eq{\alpha\equiv g^2,~~~\beta\equiv\frac{g^2}{6\pi^2Q^4\eta}.}{lab}
Then RG equations turn into
\eq{\dpf{\alpha}{l}=\alpha^2\fkh{\xi_3-\frac{3}{2\pi^2}},}{lab}
\eq{\dpf{\beta}{l}=\alpha\beta(\frac{2}{3}\xi_1-\xi_3)+\frac{4}{3}\beta-\beta^2,}{lab}
where $ \xi_i $ are dependent on $ x=\frac{6\pi^2\beta}{\alpha} $. By numerically iterating the above flow equations, we can show the stable fixed point is $ (\alpha,\beta)=(0,\frac{4}{3}) $ beyond the unstable Gaussian fixed point $ (\alpha,\beta)=(0,0) $. This results is consistent with RG flow directly calculated by $ g $ and $ \eta $.

The asymptotic behavior for $ \xi_i $ when $ x $ is around zero, i.e. near the stable fixed point $ (\alpha,\beta)=(0,\frac{4}{3}) $ are
\eq{\xi_1\approx 0.013 \ln \alpha +0.002,~~~~~\xi_3\approx -0.025.}{lab}
Thus, around the stable nontrivial fixed point, the RG behavior is captured by
\eq{\dpf{\alpha}{l}\approx -0.18\alpha^2,~~~~~\dpf{\beta}{l}\approx\beta(\frac{4}{3}-\beta+0.026\alpha),}{lab}
where $ \alpha $ is marginally irrelevant and $ \beta $ is irrelevant. Note at the stable fixed point,  $\alpha=0$ guarantee the scaling properties of spatial coordinates are the same as the tree-level results $z_1=\frac{1}{3},z_3=1$.

\subsection{4. Specific heat for interacting triple-Weyl fermions}
We investigate how temperature dependence of specific heat  is affected by Coulomb interactions by scaling argument.  It is clear that free energy density scales as $f=b^{2z_1+z_3+1} f_0$, where $b= \frac{T}{\Lambda}$ and $z_1,z_3$ are scaling dimension of spatial coordinates:  $z_1=-0.004\alpha\ln\alpha-0.0007\alpha+0.33 $, $ z_3=1+0.025\alpha $ near the stable fixed point. Consequently, specific heat $C=-T\frac{\partial^2 f}{\partial T^2}$ scales as
\bea
C(T)= b^{2z_1+z_3} C_0(\Lambda),
\eea
Namely,  the flow equation of specific heat is:
\bea
\frac{\partial C(l)}{\partial l}= (2z_1+z_3)C(l) \label{diff_specific_heat} ,
\eea
where $l \equiv -\ln b$ is the flowing parameter. And since $ \alpha $ is marginal in tree level, it is more important than $ \beta $ around the fixed point and we can safely take $ \beta=4/3 $. Then
\eq{\dpf{\ln C}{l}\approx \frac{5}{3}+0.024\alpha-0.008 \alpha\ln \alpha.}{lab}

As beta function for $ \alpha $ is
$\dpf{\alpha}{l}=-0.18\alpha^2$,
we can get $ \alpha(l)=\frac{\alpha_0}{1+0.18\alpha_0 l} $. Plug this into RG equation for $ C $ and integrate $ l=-\ln b $ on both sides, we have
\al{-\ln C&\propto \frac{5}{3}\ln \frac{T_0}{T}+0.133 \ln (\frac{1+0.18\alpha_0 \ln T_0}{1+0.18\alpha_0 \ln T})+0.022(\ln(\frac{\alpha_0}{1+0.18\alpha_0\ln T_0}))^2-0.022(\ln(\frac{\alpha_0}{1+0.18 \alpha_0\ln T}))^2.}
Namely
\eq{C\propto T^{\frac{5}{3}}(\frac{1+0.18\alpha_0 \ln T_0}{1+0.18\alpha_0 \ln T})^{-0.133 }\times\me^{0.022(\ln(\frac{\alpha_0}{1+0.18 \alpha_0\ln T}))^2-0.022(\ln(\frac{\alpha_0}{1+0.18\alpha_0\ln T_0}))^2}.}{lab}
In the main text, we define $ c_1=0.18 $, $ c_2=0.133 $ and $ c_3=0.022 $ as three positive constants for simplicity. The $ T^{\frac{5}{3}} $ behavior is due to the density of states is $ \rho(\epsilon)\propto\epsilon^{2/3} $ for free triple-Weyl fermions. 
\ipic{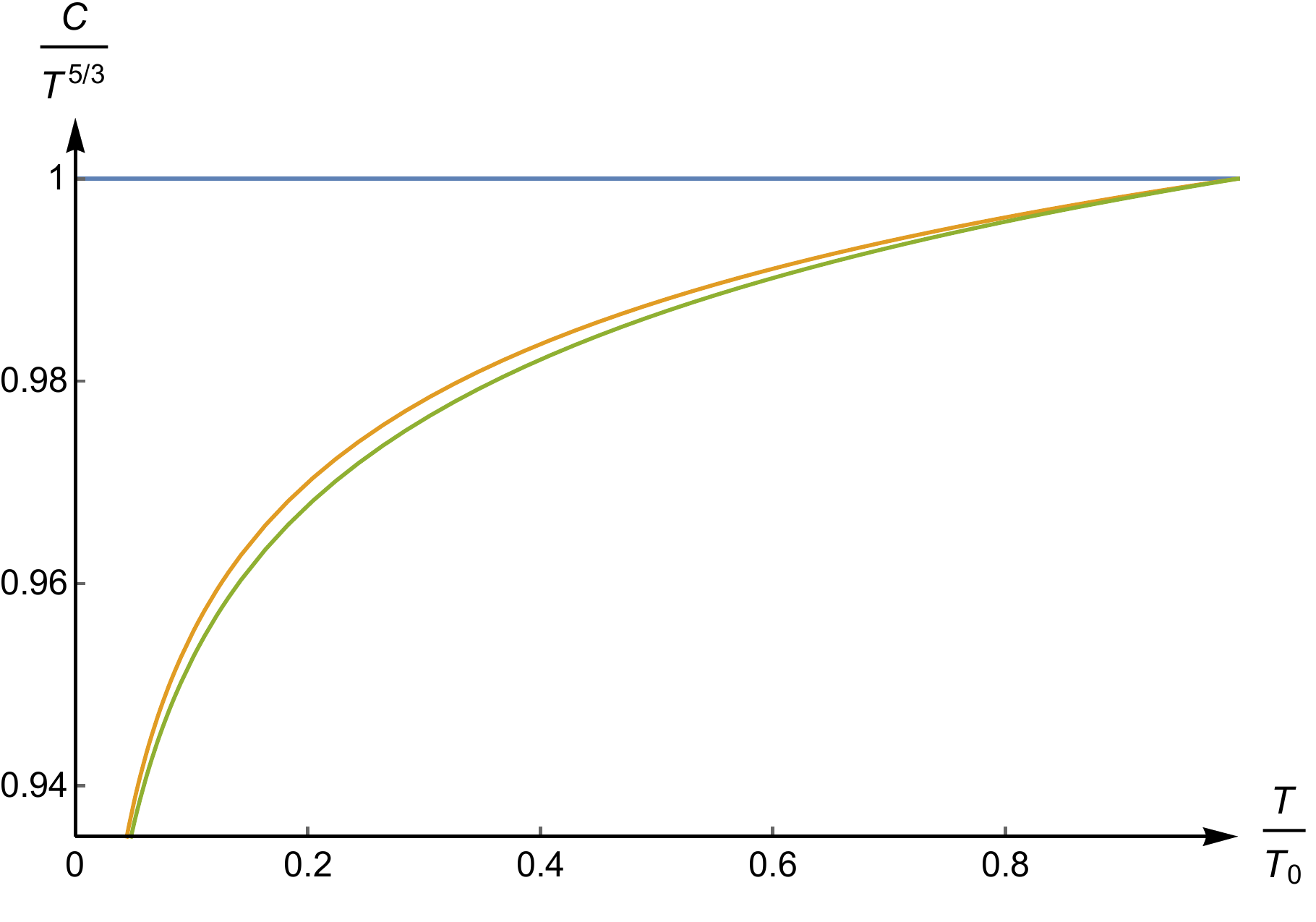}{Schematic plot of specific heat with temperature: green line is for specific heat in our case and blue one is for free triple-weyl fermions with orange one for specific heat without the exponential term}{lab}{width=10cm}

\subsection{5. Screening charge distribution induced by charge impurities}
Suppose we put some charge impurity located at origin with electric charge $ Ze $. We consider the case of noninteracting triple-Weyl fermions first. Polarization can be written as $ \Pi(\vect{q})=B_\perp q_\perp^2+B_3 q_z^{2/3} $. Induced charge density in momentum space is given by
	\begin{eqnarray}
	\rho(\vect{q}) &=& Ze V(\vect{q})\Pi(\omega=0,\vect{q})\nn\\
    &=&-Ze g_0^2 \frac{B_\perp q_\perp^2+B_3 q_z^{2/3}}{q_\perp^2+q_z^2}.
	\end{eqnarray}
We obtain $\rho(\vect{r})=\int \frac{d^3 q}{(2\pi)^3} \rho(\vect{q}) e^{-i \vect{q}\cdot \vect{r}}$ by performing  Fourier transformation,  which can be decoupled as $\rho(\vect{ r})=\rho_\textrm{I}(\vect{r})+\rho_\textrm{II}(\vect{r})$,
	\begin{eqnarray}
	\rho_\textrm{I}(r_\perp,r_z) &=& -Ze g_0^2B_\perp\frac{1}{(2\pi)^3} \int \md^2 q_\perp \md q_z \frac{
    q_\perp^2}{q_\perp^2+q_z^2} e^{-iq_z r_z-i q_\perp r_\perp} \nn\\
    &=&-Ze g_0^2B_\perp\frac{\pi}{(2\pi)^3}  \int \md^2 q_\perp |q_\perp|  e^{-|r_z| |q_\perp|} e^{-i q_\perp r_\perp} \nonumber \\
	&=& -Ze g_0^2B_\perp\frac{\pi}{(2\pi)^3}  \int \md q_\perp \md\theta q_\perp^2 e^{-|r_z| q_\perp} e^{-i q_\perp
    |r_\perp| \cos\theta} \nn\\
	&=& -Ze g_0^2B_\perp\frac{1}{4\pi} \frac{-r_\perp^2+2r_z^2}{(r_\perp^2+r_z^2)^{5/2}},
	\end{eqnarray}
and
	\begin{eqnarray}
	\rho_\textrm{II}(r_\perp,r_z) &=& -Ze g_0^2B_3 \frac{1}{(2\pi)^3} \int \md^2 q_\perp \md q_z \frac{
    q_z^{2/3}}{q_\perp^2+q_z^2} e^{-iq_z r_z-i q_\perp r_\perp}\nn\\
    &=&-Ze g_0^2B_3 \frac{1}{(2\pi)^3} \int \md^2 q_\perp  \fkh{2\pi q_\perp^{2/3} \cosh q_\perp r_z-\frac{4}{\sqrt{3}} q_\perp r_z^{1/3}\Gamma (-\frac{4}{3}){}_1F_2(1;\frac{2}{3},\frac{7}{6};\frac{q_\perp^2 r_z^2}{4})}e^{-i q_\perp r_\perp}   \nn\\
	&=& -Ze g_0^2B_3 \frac{1}{2^{1/3}\pi} \frac{\Gamma(\frac{5}{6}){}_2F_1(\frac{5}{6},\frac{5}{6};\re{2};-\frac{r_z^2}{r_\perp^2})}{r_\perp^{5/3}\Gamma(\re{6})},\nn\\
	\end{eqnarray}
	where $ {}_2F_1 $ is hypergeometric $ {}_2F_1 $ function.

Finally we obtain total screening charge density in real space:
	\begin{eqnarray}
	\rho(\vect{r}) &=& \rho_\textrm{I}(\vect{r})+\rho_\textrm{II}(\vect{r})\nn\\
    &=& -\frac{Ze g_0^2}{\pi} \left(\frac{B_\perp}{4} \frac{-r_\perp^2+2r_z^2}{(r_\perp^2+r_z^2)^{5/2}}+ \frac{B_3}{2^{1/3}}\frac{\Gamma(\frac{5}{6}){}_2F_1(\frac{5}{6},\frac{5}{6};\re{2};-\frac{r_z^2}{r_\perp^2})}{r_\perp^{5/3}\Gamma(\re{6})}\right).
	\end{eqnarray}
The screening charge distribution in real space is shown as Figure \ref{IC} in the main text.

As for the interacting triple-Weyl fermions, where Coulomb interactions become $ V(\vect{q})=\frac{g^2}{q_x^2+q_y^2+q_z^{2/3}} $, (momentum are rescaled for simplicity). Then the induced charge density in momentum space are 
\eq{\rho(\vect{q})=-Zeg^2\frac{B_\perp q_\perp^2+B_z q_z^{2/3}}{q_\perp^2+q_z^{2/3}}.}{lab}

We can calculate the total charge density for each $ r_\perp $ or $ r_z $ as
\al{Q_\perp (r_\perp)=-Zeg^2\int\md r_3\int\frac{\md^3 q}{(2\pi)^3}\frac{B_\perp q_\perp^2+B_z q_z^{2/3}}{q_\perp^2+q_z^{2/3}}\me^{-\mi \vect{q}\cdot\vect{r}}=-Zeg^2B_\perp\delta^2(r_\perp),\\Q_z (r_z)=-Zeg^2\int\md^2 r_\perp\int\frac{\md^3 q}{(2\pi)^3}\frac{B_\perp q_\perp^2+B_z q_z^{2/3}}{q_\perp^2+q_z^{2/3}}\me^{-\mi \vect{q}\cdot\vect{r}}=-Zeg^2B_z\delta(r_z).}

We obtain the above results by making full use of properties for delta functions and we further employ the requirements $Q_{\text{total}}=\int\md^2 r_\perp Q_\perp (r_\perp)=\int\md z Q_z(r_z)  $ to get the relation $ B_\perp=B_z $. Finally, we conclude that induced charges in interacting case distribute as $ \rho(\vect{q})=\text{const} $ and $ \rho(\vect{r})\propto\delta^{(3)}(\vect{r}) $ in real space which is located around the charge impurity.

\end{widetext}

\end{document}